\documentclass[12pt]{article}

\usepackage{scicite}

\usepackage{txfonts, graphicx}

\def\aj{AJ}                   
\def\apj{ApJ}                 
\def\apjs{ApJS}               
\def\aaps{A\&AS}              
\def\mnras{MNRAS}             
\def\pasp{PASP}               
\def\nat{Nature}              
\def\iaucirc{IAU~Circ.}       

\newenvironment{sciabstract}{%
\begin{quote} \bf}
{\end{quote}}

\renewcommand\refname{References and Notes}

\newcounter{lastnote}

\title{The Disappearance of the Progenitors of Supernovae 1993J and 2003gd}

\author
{Justyn  R. Maund,$^{\ast,1,2,3}$ Stephen J. Smartt,$^{4}$\\
\\
\normalsize{$^{1}$ Dark Cosmology Centre, Niels Bohr Institute, University of Copenhagen}\\
\normalsize{Juliane Maries Vej 30, 2100 Copenhagen \O, Denmark}\\
\normalsize{$^{2}$ Department of Astronomy \& Astrophysics, University of California, Santa Cruz, 95064, U.S.A.}\\
\normalsize{$^{3}$ Sophie \& Tycho Brahe Fellow}\\
\normalsize{$^{4}$ Astrophysics Research Centre, School of Mathematics and Physics,}\\
\normalsize{Queens' University Belfast, Belfast, BT7 1NN, United Kingdom}\\
\\
\normalsize{$^\ast$To whom correspondence should be addressed;}\\
\normalsize{E-mail:  justyn@dark-cosmology.dk.}
}

\date{}

\begin{document} 


\baselineskip24pt


\maketitle


\begin{sciabstract}
  Using images from the Hubble Space Telescope (HST) and the Gemini
  Telescope we confirm the disappearance of the progenitors of two
  Type II supernovae (SNe), and evaluate the presence of other stars
  associated with them.  We find that the progenitor of SN~2003gd, an
  M-supergiant star, is no longer observed at the SN location, and
  determine its intrinsic brightness using image subtraction techniques.
  The progenitor of SN~1993J, a K-supergiant star, is also no longer
  present, but its B-supergiant binary companion is still observed.
  The disappearance of the progenitors confirms that these two SNe were
  produced by Red Supergiants.
\end{sciabstract}

Analysis of archival pre-explosion images has helped to identify the
progenitors of type II supernovae, which are thought to result from
the explosion of red supergiant stars.  In most cases the progenitors
have only been confirmed to be spatially coincident with the SNe,
leaving some uncertainty over their correct identification.  So far
only one star has been shown to have disappeared after it exploded -
the star that exploded as SN~1987A in the Local Group of galaxies
\cite{gil87a}. Seven other stars have been discovered to be spatially
coincident with bright, nearby type II SNe \cite{2008arXiv0809.0403S},
but none of them have been shown to have disappeared [although recent
evidence has suggested the disappearance of the progenitor of SN2005gl
\cite{galyamnote}].  Here we examine post-explosion images of the
sites of two type II SNe to confirm their progenitors'
identity.\\

SN~2003gd was discovered in the galaxy M74 and classified as a type
II-plateau SN \cite{2003IAUC.8150....2E}.  The plateau in its light
curve and the strong hydrogen P Cygni profiles in its spectrum
indicated that it arose from a red supergiant star with a massive
hydrogen envelope \cite{2005MNRAS.359..906H}.  Analysis of
pre-explosion images from the Hubble Space Telescope (HST) and Gemini
Telescope archives revealed a red supergiant star, with $V=25.8$,
$I=23.3$ and an initial mass of $8^{+4}_{-2}$ solar masses ($M_{\odot}$), at the
position where the SN occurred (see Fig. 1;
\cite{smartt03gd,2003PASP..115.1289V}). Five years after the explosion
it is time to verify that this star has disappeared, hoping that the
SN remnant has faded to a luminosity below that of its progenitor (as
achieved for SN~1987A).\\

We re-imaged the site of SN~2003gd on 6 and 10 September 2008 using
the multiobject spectrograph [GMOS, \cite{2004PASP..116..425H}] on the
Gemini Telescope with the Sloan $g'$, $r'$ and $i'$ filters, under
excellent seeing conditions (see Table 1).  Using differential
astrometry, we matched previous observations of SN~2003gd, acquired
with the HST Advanced Camera for Surveys High Resolution Channel (ACS
HRC)\cite{smartt03gd}, with our GMOS frames to locate the SN position
to within $0.031"$ (see Fig. 1).  At this position there is no
significant detection of a point source in the $i'$ image, but the SN
is still detected significantly at $m_{g'}(AB)=25.00\pm0.04$ and
$m_{r'}(AB)=24.65\pm0.05$ (see SOM).  We compared our $i'$ image with
a pre-explosion $i'$ image taken with the GMOS instrument on 14 Aug
2001, in which the progenitor candidate was detected.  After aligning
the two images, we scaled the flux levels and the Point-Spread
Function (PSF) of the post-explosion $i'$-band image to match those of
the pre-explosion image, using the $ISIS$ image subtraction package
\cite{1998ApJ...503..325A,2000A&AS..144..363A}.  We then subtracted
the scaled post-explosion image from the pre-explosion frame, such
that only photometrically variable objects remained (see the SOM for
details of the procedure).  A point source residual, consistent with a
single-star PSF, is present in the pre-explosion image to within
$0.023''$ of the transformed SN position (see Fig. 1, see SOM).
Ref. \cite{smartt03gd} found a large offset between the transformed SN
position and the object on the pre-explosion frame ($0.137\pm0.071"$),
leading them to suggest that the object was a blend of the progenitor
(Star A) and a nearby star (Star C, see Fig. 1).  We find no such
offset with our differential astrometric solution, such that the
object on the pre-explosion frame is completely consistent with a
single star located at the position of Star A.  All other single
non-variable stellar objects were cleanly subtracted from the
pre-explosion image, with only minor residuals remaining for obviously
extended features such as resolved clusters.  Hence the point source
visible in the subtracted frame is the $i'$-band flux from the
progenitor which has now disappeared. There clearly is a faint,
extended background feature in the post-explosion image, but this is
inconsistent with a single PSF.  By conducting forced photometry at
the SN position and by adding artificial stars to our post-explosion
$i'$ image, and attempting to recover them using the $ISIS$ package,
we conservatively estimate the upper brightness limit for a single
point source is $m_{i}(AB) > 26.3$ at the SN location.  This
corresponds to a magnitude difference between the pre- and
post-explosion $i'$ images of $\Delta m_{i'} > 2$. The absence of
an obvious point source in the post-explosion image, which is within 2
magnitudes of the detected progenitor, further
confirms that the star has disappeared.\\

We transformed the pre-explosion $i'$ band photometry to the Johnson
system by bootstrapping the photometry with Wide Field Camera images,
from the Isaac Newton Telescope, and the M74 standard star sequence
used for monitoring both SN~2003gd and SN~2002ap
\cite{2002GCN..1242....1H}.  These data imply that the progenitor star
had a final Johnson $I$ band magnitude of $23.14\pm0.08$, tightening
earlier constraints \cite{smartt03gd}.   This is 0.2 mags brighter than previously reported, due to our determination that Star C contributed negligible $i'$ flux.\\

Using the HSTphot package \cite{dolphhstphot}, we analyzed
pre-explosion HST WFPC2 images taken with the $F606W$ filter, which is
close to the V-band filter.  We used the $I$-band magnitude determined
above to solve the color-transformation equations that convert the
$F606W$ instrumental magnitude to the Johnson system, and found that
the progenitor has a Johnson V magnitude of $25.72\pm0.09$.
Correcting for a reddening of $E(B-V)=0.14\pm0.06$, which was
determined from photometry of SN~2003gd \cite{2005MNRAS.359..906H},
implies an intrinsic color of $(V-I)_{0}=2.41\pm0.14$.  This is
consistent with the progenitor being a M0-M2 Red Supergiant star
\cite{1985ApJS...57...91E}, but disfavors warmer K-supergiant stars.
Applying the appropriate bolometric correction for this range of
allowed red supergiants, and correcting for a distance of
$\mathrm{9.3\pm1.8\,Mpc}$ to M74 \cite{2005MNRAS.359..906H}, yields a
luminosity of $\mathrm{log}(L/L_{\odot})=4.29\pm0.2$ with an effective
temperature of $\mathrm{log}T_{eff}=3.54\pm0.02$.  Comparison with the
end-points of stellar evolution models shows that this luminosity and
temperature region of the Hertzsprung-Russell diagram constrains the
progenitor to have had initial mass of $7^{+6}_{-1}M_{\odot}$
\cite{2008arXiv0809.0403S}.\\

At the epoch of 6 September 2008 (approximately $+2000$ days after
explosion), the SN is still visible in the wavelength range
corresponding to the F622W and Sloan $r'$ filters (See Table S7).
This is due to the presence of significant $H\alpha$ emission in the
band-passes of these filters, typical of Type II SNe at late times
\cite{2005MNRAS.359..906H,andrea05cs}.  The fact that the SN is bright
in the late-time F622W observations precludes the use of image
subtraction techniques to determine the true progenitor brightness in
the pre-explosion WFPC2 F606W image.  We need to wait longer until
the $H\alpha$ flux fades well below the magnitude of the
progenitor star detected in the F606W filter.\\

One could argue that the star identified as the progenitor was a
neighbouring star that is now obscured by dust formation in the
foreground SN remnant.  However the internal extinction in the
SN~2003gd remnant, due to newly formed dust was estimated to be $A_{R}
< 1.48$ \cite{2006Sci...313..196S} or $A_{i'} <1.22$ \cite{ccm89} at
678 days.  Using this as the maximum value of the extinction across
the SN remnant at the epoch of our GMOS images (assuming no further
dust formation), this amount of extinction is insufficient to cause
$\Delta m_{i'}$ measured by us.  This implies that the object we have
identified as the progenitor is not simply obscured by dust in the intervening SN remnant and has actually disappeared.\\

The progenitor of the Hydrogen-bearing (Type IIb) SN~1993J was
identified as a K-supergiant star, with excess flux in the
ultra-violet, possibly because of a binary companion or nearby stars
\cite{alder93j}.  The model for this binary system was of a $15
M_{\odot}$ progenitor star, with a binary companion of slightly lower
mass
\cite{1993Natur.364..507N,1993Natur.364..509P,1994ApJ...429..300W}.
Because the progenitor star evolved faster, it underwent mass transfer
onto the binary companion, which removed a substantial amount of its
Hydrogen envelope, causing a shift to the bluer K-spectral type
[rather than the canonical M spectral type
\cite{2008arXiv0809.0403S}].  The binary companion grew to
$22M_{\odot}$, and became the source of the excess ultra-violet flux.
A later study, using high-resolution HST images, found that some of
the UV excess could be explained by nearby previously unresolved
stars, but that a UV excess still remained unaccounted for
\cite{2002PASP..114.1322V}.  The binary progenitor scenario was
confirmed, when spectral features of a massive blue supergiant were
detected in late-time
spectroscopy \cite{maund93j}.\\

The site of SN~1993J was imaged several times over the 2-13 years
after explosion with the HST WFPC2, ACS HRC and Wide-field Channels (WFC)
(see SOM for full list of dates).  By the epoch of the 2004
observation, the red portion of the SN spectral energy distribution
(SED) had faded below the level of the SED of the binary progenitor
system (Fig. 2, see SOM), ruling out the continued presence of the
K-supergiant star and, hence, confirming it as the progenitor of
SN~1993J.  At the current rate of decay of the SN SED, the U and
B-band fluxes will have reached the level of the proposed B-supergiant
component by 2012, at which stage it will be possible to directly
measure the properties of the
remaining companion star.\\

These results provide observational proof that red supergiant stars
are the progenitors of type II SNe, through the disappearance of the
previously identified candidate stars of two SNe.  Our best estimate
for the mass of the progenitor of the Type IIP SN~2003gd is
$7M_{\odot}$ which is at the lower end of the mass range considered
theoretically possible to produce core-collapse events.  While the
uncertainties ($7^{+6}_{-1}M_{\odot}$) would comfortably allow a large
mass for this object, it is interesting that five progenitors of Type
IIP SNe are found with best estimates at $9M_{\odot}$ or below
\cite{2008arXiv0809.0403S}.  This limit is predicted by stellar and SN
evolution models \cite{eld04} and is consistent with the upper initial
mass limit observed for white dwarfs \cite{2008arXiv0811.1577S}.  The
confirmation of the disappearance of the K-type progenitor star of
SN1993J is further evidence that the binary model previously suggested
is valid. It demonstrates the importance of binary interactions for
the production of Hydrogen
poor SNe \cite{2004ApJ...612.1044P}.\\

\clearpage

\begin{figure}
\includegraphics[width=16cm]{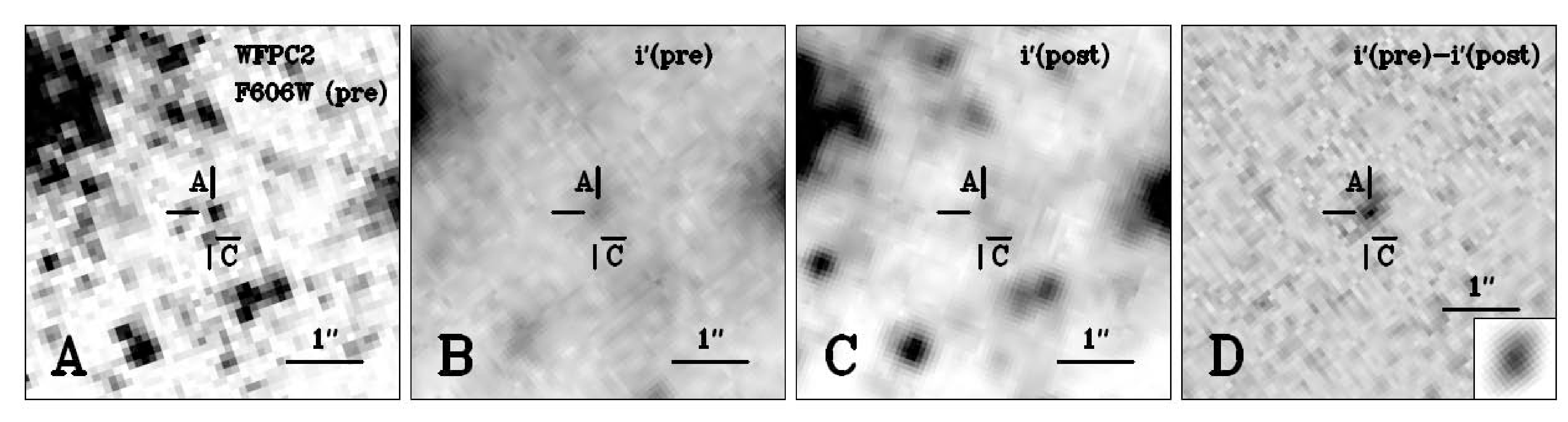}
\caption{Pre- and post-explosion images of the site of SN~2003gd taken
  with the HST WFPC2 and Gemini GMOS instruments (see Table 1 for
  details).  {\bf (A)} Pre-explosion WFPC2 F606W image, with the
  previously identified progenitor object marked as Star A, and a
  nearby star labelled Star C.  {\bf (B)} Pre-explosion Gemini GMOS
  $i'$ band image, with spatial resolution $0.57"$, in which the
  progenitor is detected. {\bf (C)} Post-explosion Gemini GMOS $i'$
  band image where a single point source is not detected at the
  transformed position of SN~2003gd, with a limit on any remaining SN
  flux of $m_{i'}(AB) > 26.3$. The image has a spatial resolution of
  $0.36"$ and no point source is detected at the position of Star C,
  suggesting a negligible contribution in the $i'$-band from that
  star.  {\bf (D)} Pre-explosion image with the flux/PSF-scaled
  post-explosion image subtracted.  The residual object at the
  position of SN~2003gd is the progenitor, with any contaminating flux
  from nearby stars subtracted.  The object at the SN location, marked
  as Star A, is consistent with single-star PSF ($\chi^{2}=0.9$) and
  has a Johnson-I magnitude $23.14\pm0.08$
  ($m_{i'}(AB)=24.25\pm0.04$).  In the bottom right hand corner a
  model single star PSF is shown.  It was determined by $ISIS$ and is
  consistent with the source detected in this subtraction image.}
\end{figure}

\clearpage
\begin{figure}
\rotatebox{-90}{\includegraphics[width=6cm]{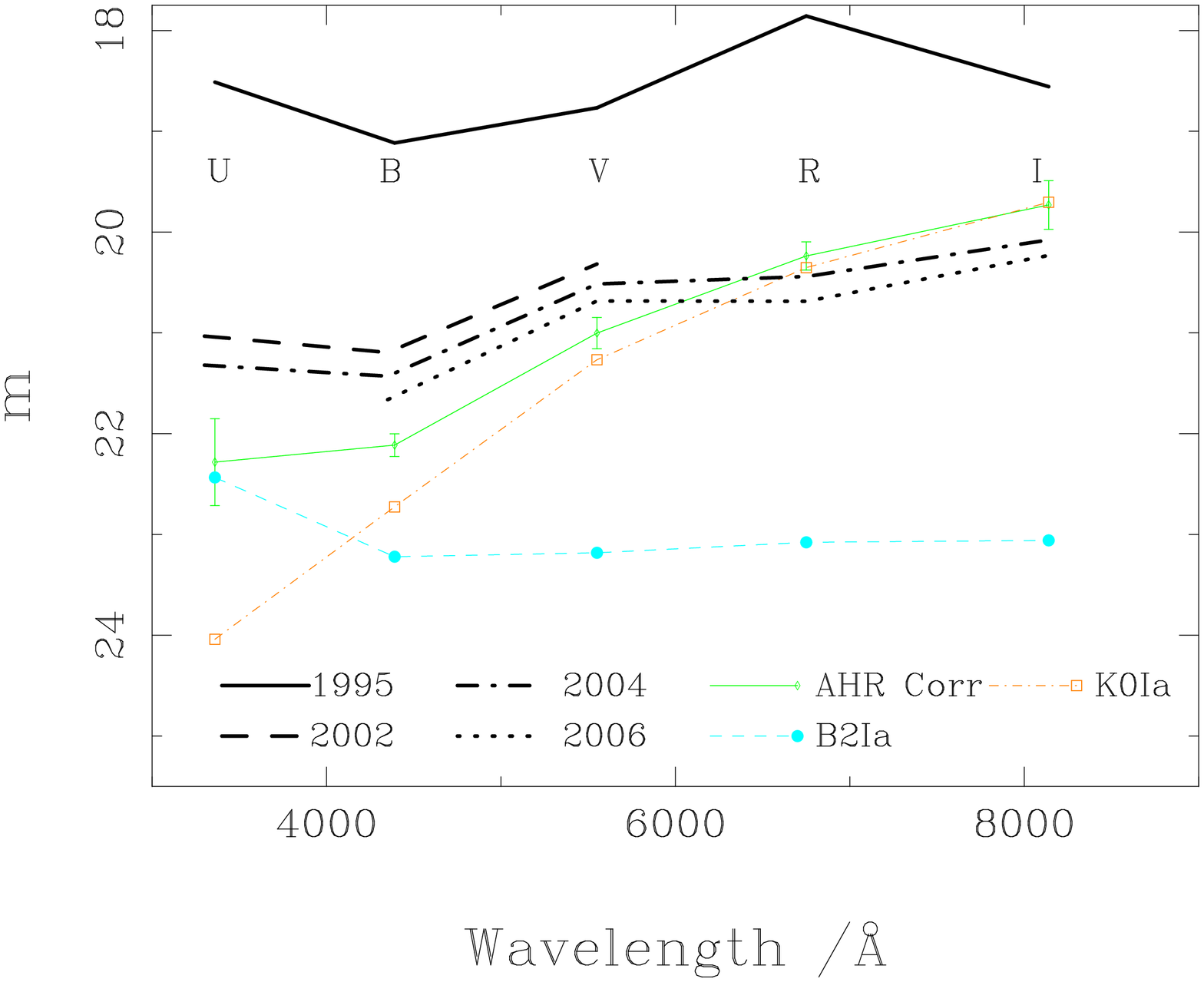}}\rotatebox{-90}{\includegraphics[width=6cm]{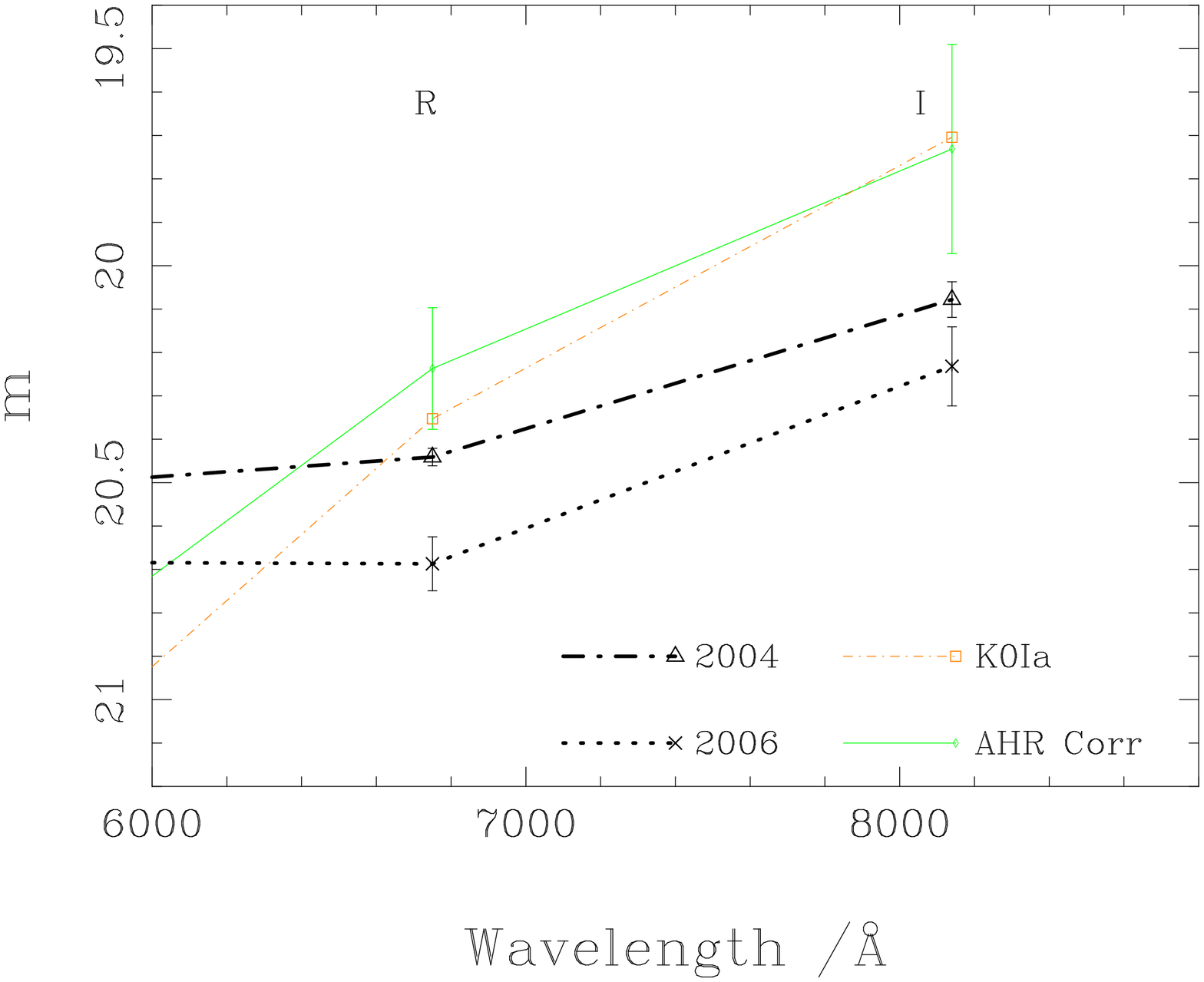}}
\caption{The Spectral Energy Distribution of the progenitor binary system of
  SN~1993J.  The SED (AHR Corr)  has been corrected for the excess flux
  contribution from four nearby stars, unresolved in the
  low-resolution ground-based pre-explosion imaging, for comparison
  with photometry of high-resolution HST imaging in which the SN is
  resolved. {\bf (Left)} Overlaid SN~1993J
  SEDs, measured with HST WFPC2, ACS/HRC and ACS/WFC, at 31 Jan 1995,
  28 May 2002, 15 Jul 2004 and 1 and 3 Nov 2006.  The B and K
  supergiant components are the result of a $\chi^{2}$ fit to the
  excess corrected pre-explosion photometry \cite{maund93j,mythesis}.
  {\bf (Right)} A zoom in on the $R_{C}$ and $I_{C}$ photometric
  points.  Errors on the 2004 and 2006 photometry are calculated from
  PSF photometry of SN~1993J.}
\end{figure}

\clearpage
\begin{table}
\caption{Pre- and post-explosion observations of the site of SN~2003gd.  (*pixel scale)}
\begin{tabular}{llcccc}
Telescope/      & Date      &  Program    &     Filter      &   Exposure    &  Seeing     \\
Instrument      &           &              &                 &  Time(s)      &   ($"$)     \\
\hline
Gemini GMOS     & 14 Aug 2001 & GN-2001B-SV-102 &       $i'$      & 480      &   0.57      \\
HST/WFPC2       &25/28 Aug 2002 & GO-9676 &    F606W        &    2100       &   0.1*          \\
\\
HST/ACS/HRC     & 01 Aug 2003 & GO-9733  &    F814W        & 1350           &  0.025*           \\
\\
Gemini GMOS     &10 Sep 2008&GN-2008B-Q-67&      $g'$       &  4050         &  0.53       \\
Gemini GMOS     &06 Sep 2008&GN-2008B-Q-67&      $r'$       &  1590         &  0.42       \\
Gemini GMOS     &06 Sep 2008&GN-2008B-Q-67&      $i'$       &  3180         &  0.36       \\
\end{tabular}
\end{table}

\clearpage
\bibliographystyle{Science}

27. The authors thank C. Alard for discussion about the $ISIS$
  code and Joanna Fabbri for discussion concerning dust formation in
  the remnant of SN~2003gd.  The research of JRM is funded through the
  Sophie \& Tycho Brahe Fellowship.  The Dark Cosmology Centre is
  supported by the DNRF.  This work, conducted as part of the award
  "Understanding the lives of massive stars from birth to supernovae"
  made under the European Heads of Research Councils and European
  Science Foundation EURYI Awards scheme, was supported by funds from
  the Participating Organisations of EURYI and the EC Sixth Framework
  Programme.  Based on observations made with the NASA/ESA Hubble
  Space Telescope, obtained from the Data Archive at the Space
  Telescope Science Institute, which is operated by the Association of
  Universities for Research in Astronomy, Inc., under NASA contract
  NAS 5-26555. These observations are associated with programs 6139,
  9353, 9676, 9733, 10204, 10250, 10272, 10584, 10877, and 11229.
  Based on observations obtained at the Gemini Observatory, which is
  operated by the Association of Universities for Research in
  Astronomy, Inc., under a cooperative agreement with the NSF on
  behalf of the Gemini partnership: the National Science Foundation
  (United States), the Science and Technology Facilities Council
  (United Kingdom), the National Research Council (Canada), CONICYT
  (Chile), the Australian Research Council (Australia), Minist\'{e}rio
  da Ci\^{e}ncia e Tecnologia (Brazil) and SECYT (Argentina).  Gemini
  data acquired for programs GN-2001B-SV-102 and GN-2008B-Q-67.

\clearpage
\begin{centering}
{\bf\Large Supporting Online Material: The Disappearance of the Progenitors of Supernovae 1993J and 2003gd}
\end{centering}
\renewcommand{\thefigure}{S\arabic{figure}}
\renewcommand{\thetable}{S\arabic{table}}
\renewcommand{\thesection}{S\arabic{section}}
\renewcommand\refname{References and Notes}
\setcounter{table}{0}
\section{Observations}
The observations used in this study are listed in Tables \ref{93JACS}, \ref{93JWFPC2} for SN~1993J and Tables \ref{03gdACS}, \ref{03gdWFPC2}, \ref{03gdGEM} and \ref{03gdINT} for SN~2003gd.  Due to space constraints, in the main body of the paper only the most appropriate observations were explicitly discussed.   All the observations presented in these tables were also reduced, analysed and considered as part of the broader context of the study, and corroborated the results presented here.
\begin{table}
\caption{\label{93JACS}Late-time HST ACS observations of the site of SN~1993J.}
\begin{tabular}{lllll}
Date   &  Dataset  &  Filter & Exp Time &   \\
(JD +2\,450\,000) &  & & (secs) & \\
\hline
2002 May 28$^{1}$ & J8DT07010 & F250W & 2100 & HRC \\
(2422.81)       & J8DT07020 & F330W & 1200 & HRC \\
                         & J8DT07030 & F435W & 1000 & HRC \\
                         & J8DT07040 & F555W & 1120 & HRC \\
\\
2004 Jul 15$^{2}$   & J8ZO02010 & F330W  & 2400 & HRC \\
(3203.40)       & J8ZO02040 & F435W & 1548 & HRC \\
                         & J8ZO02020 & F555W & 1200 & HRC \\
                         & J8ZO02030 & F625W & 1200 & HRC \\
                         & J8ZO02050 & F814W & 1739 & HRC \\
\\ 
2004 Sep 15$^{3}$  & J90L08010  & F814W & 1650 & WFC \\
(3262.82) \\
\\
2005 Jan 11$^{2}$  & J8ZO03010 & F625W & 1200 & HRC \\
(3382.38)        & J8ZO03020 & F814W & 1404 & HRC \\
\\
2006 Sep 08$^{4}$  & J9EL14010 & F435W & 1200 & WFC  \\
(3987.42)        & J9EL14020 & F606W & 1200 & WFC \\
\\
2006 Nov 1$^{5}$     & J9NW02010 & F555W & 480  & HRC \\
(4040.78)        & J9NW02020 & F814W &720 & HRC \\
\\
2006 Nov 3$^{5}$     & J9NW01010 & F435W & 840 & HRC \\
(4042.77)        & J9NW01020  & F625W & 360 & HRC \\
\end{tabular}
\\
$^{1}$ Program 9353 (P.I. S.J. Smartt)\\
$^{2}$ Program 10204 (P.I. B.E. Sugerman)\\
$^{3}$ Program 10250 (P.I. J. Huchra)\\
$^{4}$ Program 10584 (P.I. A. Zezas)\\
$^{5}$ Program 10877 (P.I. W. Li)\\
\end{table}

\begin{table}
\caption{\label{93JWFPC2}Late-time HST WFPC2 observations of the site of SN~1993J.}
\begin{tabular}{llll}
Date   &  Dataset  &  Filter & Exp Time \\
(JD +2\,450\,000) &  & & (secs)  \\
\hline
1995 Jan 31$^{1}$ & U2MH0101T & F255W & 2400 \\
(-251.06)                        & U2MH0105T & F336W & 1160  \\
                        & U2MH0108T & F439W &   1200 \\
                        & U2MH010BT & F555W &  900  \\
                        & U2MH010DT & F675W &  900  \\
                        & U2MH010FT & F814W &   900  \\
\\
2001 Jun 4$^{2}$ & U6EH0101R & F555W & 2000\\
(2064.69)                                 & U6EH0105R & F450W & 2000\\
                                 & U6EH0109R & F814W & 2000\\
\\
2004 Sep 25 $^{3}$ & U8ZO0101M & F675W & 1200 \\
(3273.66)                                     & U8ZO0102M & F555W & 1100 \\                    
\end{tabular}                               
\\
$^{1}$ Program 6139 (P.I. R. Kirshner)\\
$^{2}$ Program 9073 (P.I. J. Bregman )\\
$^{3}$ Program 10204 (P.I. B. Sugerman)\\
\end{table}

\begin{table}
\caption{\label{03gdACS}HST ACS observations of the site of SN~2003gd}
\begin{tabular}{lllll}
Date   &  Dataset  &  Filter & Exp Time &   \\
(JD +2\,450\,000) &  & & (secs) & \\
\hline
2003 Aug 01$^{1}$ &  J8NV01011 & F435W & 2500  & HRC \\
(2700.93)                      & J8NV01031  & F555W & 1100 & HRC \\
                         & J8NV01051 & F814W & 1350 & HRC \\
\\                      
2004 Dec 08$^{2}$ & J8Z447011 & F435W & 840 & HRC \\
(3347.93)                                      & J8Z447021 & F625W & 360 & HRC \\
                                     \\                
\end{tabular}
\\
$^{1}$ Program 9733 (P.I. S.J. Smartt)\\
$^{2}$ Program 10272 (P.I. A.V. Filippenko) \\
\end{table}

\begin{table}
\caption{\label{03gdWFPC2}HST WFPC2 observations of the site of SN~2003gd.}
\begin{tabular}{llll}
Date   &  Dataset  &  Filter & Exp Time \\
(JD +2\,450\,000) &  & & (secs)  \\
\hline
2002 Aug 28$^{1}$ & U8IXCY01M & F606W & 2100 \\
(2514.98)\\
\\
2007 Jun 21$^{2}$ & U9NW0301M & F450W & 800 \\
(4273.26)                                   & U9NW0303M & F675W & 360 \\
                                   \\
2007 Aug 11$^{3}$ & UA2P0501M & F622W & 1600 \\
(4323.61)                                    & UA2P0505M & F814W & 1600 \\

\end{tabular}
\\
$^{1}$ Program 9676 (P.I. J. Rhoads)\\
$^{2}$ Program 10877 (P.I. W. Li)\\
$^{3}$ Program 11229 (P.I. M. Meixner)
\end{table}
\begin{table}
\caption{\label{03gdGEM}Gemini GMOS-N observations of the site of SN~2003gd.}
\begin{tabular}{llll}
Date   &  Dataset  &  Filter & Exp Time   \\
(JD +2\,450\,000) &  & & (secs) \\
\hline
2001 Aug 14$^{1}$ & 1-001 & r' & $4 \times 120$ \\
(2136.07)                                    & 1-005 & i' & $4 \times 120$ \\
                                    & 1-009 & g' & $4 \times 120$\\
                                   \\
2008 Sep 06$^{2}$ & 2-002 & r' & $3 \times 530$ \\
(4716.12)                                    & 3-001 & i' & $6 \times 530$ \\                                  
\\
2008 Sep 10$^{2}$ & 1-001 & g' & $5 \times 810$\\
(4720.12)
\\
\end{tabular}
\\
$^{1}$ Program GN-2001B-SV-102 (P.I. Gemini Staff)\\
$^{2}$ Program GN-2008B-Q-67 (P.I. J.R. Maund) \\
\end{table}

\begin{table}
\caption{\label{03gdINT}INT WFC observations of the site of SN~2003gd.}

\begin{tabular}{llll}
Date   &  Dataset  &  Filter & Exp Time   \\
(JD +2\,450\,000) &  & & (secs) \\
\hline
2001 Jul 24$^{1}$ & r268285  & V & 10 \\
(2114.71) & r268286 & V & 120 \\
 & r268287 & B & 120 \\
 &r268289 & I & 120 \\
\end{tabular}
\\
$^{1}$ Observer E.P.J. van den Heuvel\\
\end{table}

The HST data were retrieved from the Space Telescope Science Institute
Archive \footnote{http://archive.stsci.edu/hst}.  These data were processed by
the on-the-fly-recalibration pipeline, being processed with the most
up-to-date calibration frames available for the WFPC2, ACS/HRC and
ACS/WFC instruments, and were retrieved in a form on which analysis
could be directly conducted.\\
The Gemini-North GMOS data were retrieved from the Gemini Science
Archive\footnote{http://archive.gemini.edu/}, and were reduced with
master bias and flatfield calibration frames acquired at the same
epochs.  The individual observations were combined using the
IRAF\footnote{IRAF is distributed by the National Optical Astronomy
  Observatory, which is operated by the Association of Universities
  for Research in Astronomy (AURA) under cooperative agreement with
  the National Science Foundation.} {\tt gemini} package {\it imcoadd}
task to produce combined master science frames, on which photometry
could be conducted.  As a dithering pattern had been employed, the
coaddition of the subimages for each filter filled in any gaps between
the detectors as well as move the location of any hot pixels relative
to observed stars.  The 2008 observations were conducted under
photometric conditions, for which photometric zeropoints (for the
Sloan AB magnitude system) were calculated in the standard
way\footnote{http://www.gemini.edu/sciops/instruments/gmos/calibration/photometric-stds}.\\
Fully reduced observations of M74 using the Isaac Newton Telescope
Wide field Camera, with the Harris $BVI$ filters, were retrieved from
the Cambridge Astronomical Survey Unit archive
\footnote{http://archive.ast.cam.ac.uk}.\\

\section{Photometry}
\subsection{HST data}

Photometry of the HST WFPC2 data was conducted using the {\sc HSTphot}
package
\cite{dolphhstphot}\footnote{http://purcell.as.arizona.edu/hstphot/},
which provides corrections for aperture size and charge transfer
efficiency, and includes transformations for photometry from the WFPC2
magnitude system to the standard Johnson-Cousins system.  In addition,
in parallel, our own photometry was conducted using the {\tt daophot}
package \cite{stet87} in IRAF.  Corrections for the charge transfer
efficiency were adopted from \cite{dolp00cte} and aperture corrections
were taken from \cite{holsper95}.  The results from this parallel
analysis were identical to those provided by {\sc HSTphot}.\\

HST ACS photometry was conducted using our own scripts (a modified
version of {\tt daophot}, which uses pre-calculated
TinyTim\footnote{http://www.stsci.edu/software/tinytim/tinytim.html}
PSFs) and the {\sc DOLphot} program
\footnote{http://purcell.as.arizona.edu/dolphot/}.  Both of these
codes provide similar corrections and transformations for the ACS
system, as {\sc HSTphot} does for the WFPC2 instrument.  {\sc DOLphot}
conducts photometry of cosmic-ray rejected, distorted, combined
``crj'' images and individual distorted ``flt'' images.  Our script
measures photometry using the distortion corrected drizzled ``drz'' images, with
the sky background unsubtracted during the drizzling process.
Importantly, the photometry using the different images and different
codes generally agreed with each other to within the photometric
errors, but for consistency we report the magnitudes reported by
the {\sc DOLphot} package.\\

\subsection{Gemini Data}
Photometry of the Gemini GMOS data was conducted in the standard
fashion using {\tt daophot}.  Aperture corrections were determined for
each of the images using bright isolated stars present in the field.
PSF models were constructed for each of the images, and photometry was
conducted by fitting PSFs to stars.  The observations were flux
calibrated using observations of the standard fields SA110-361 and
SA95-100, for the nights of 06 Sep 2008 and 10 Sep 2008 respectively.
Zeropoints were derived in the standard manner \cite{gmosphot}:
$m_{g'}=27.852\pm0.021$, $m_{r'}=28.229\pm0.037$ and
$m_{i'}=28.215\pm0.023$.\\
In parallel the {\it g'r'i'} magnitudes were transformed to the
standard Johnson-Cousins system, by bootstrapping the photometry via
INT WFC Harris BVI observations and the Henden Johnson-Cousins
standard star sequence \cite{2002GCN..1242....1H}.  The correction
from Cousins I to Johnson I magnitudes, with a quadratic dependence on
{\it V-I} color, was computed using the IRAF STSDAS {\tt synphot} package
and spectra from the
Bruzual-Persson-Gunn-Stryker Spectrophotometry Atlas.\\
To test for consistency between our bootstrapped photometry and the
absolute photometry derived from our zeropoints, we determined a
relationship between $i'(AB)-I_{c}(Vega)$ and $V-I$ using {\tt
  synphot}. This yielded a correction of $i'-I_{c}=+0.7$ at $V-I=2.5$
which is consistent with the $I_{c}$ and $i'(AB)$ magnitudes
determined for progenitor of SN~2003gd using the two separate
calibrations.

\section{SN~2003gd Images}
\begin{figure}
\hfil~\includegraphics[width=11cm]{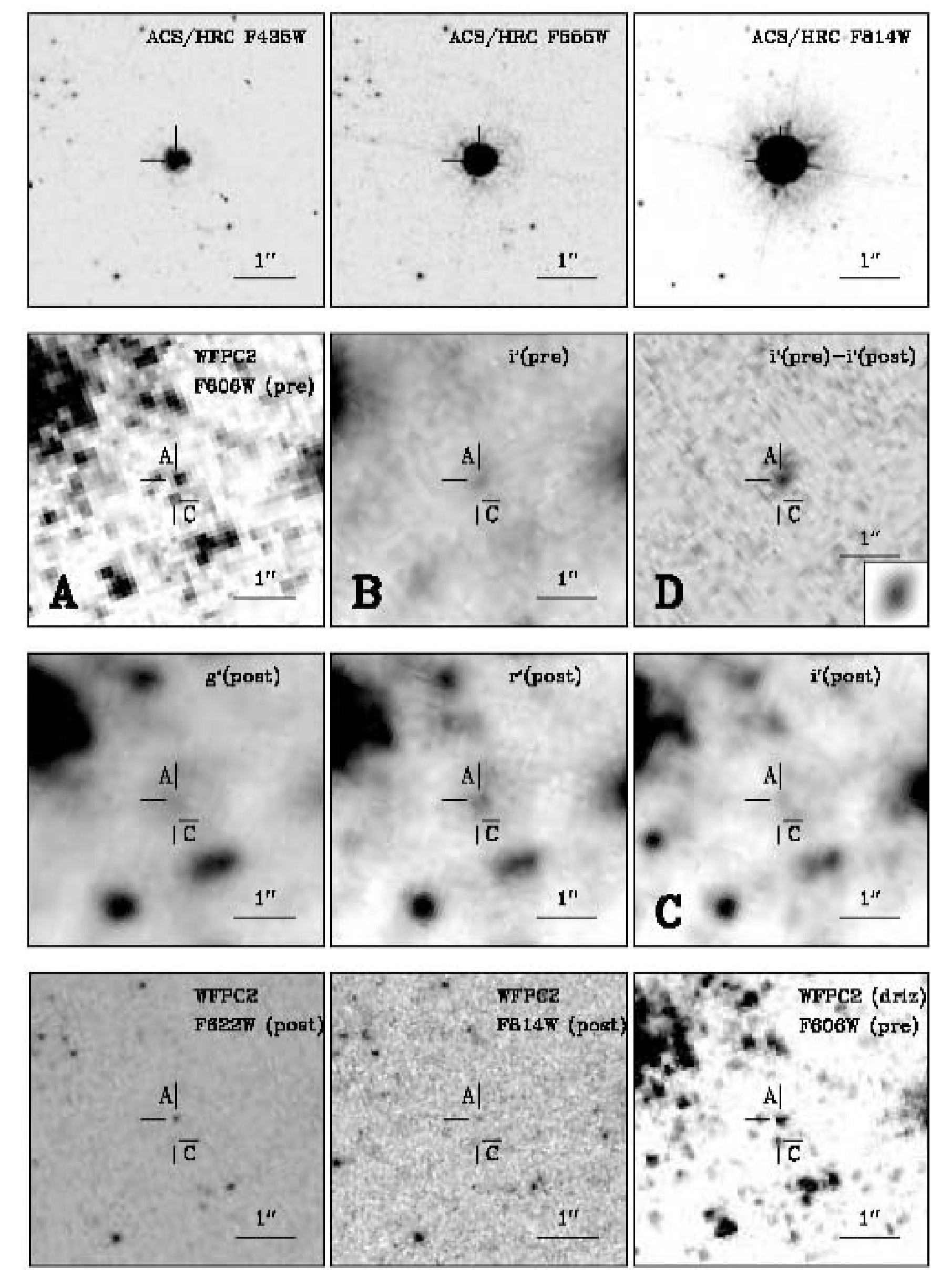}~\hfil
\caption{The site of SN~2003gd: {\bf (First Row, l to r)} Post-explosion ACS HRC/F435W, ACS/HRC F555W and ACS/HRC F814W images (2003 Aug 01). {\bf (Second Row, l to r)} Pre-explosion WFPC2/WF2 F606W image (2002 Aug 28), with the progenitor indicated as star A, pre-explosion Gemini GMOS $i'$ image (2001 Aug 14), and the {\it ISIS} subtraction image $i'$(pre)-$i'$(post).  {\bf (Third Row, l to r)} Post-explosion Gemini GMOS $g'$, $r'$ and $i'$ images (2008 Sep 10 and 2008 Sep 06). {\bf (Fourth Row, l to r)}, Post-explosion WFPC2/PC F622W and F814W images (2007 Aug 11) and dithered, drizzled pre-explosion F606W image with a pixel scale of $0.05"$.  The positions of Stars A and C, as identified by Smartt et al., \cite{smartt03gd}, are indicated by the cross-hairs.}
\label{allimages}
\end{figure}

A montage of images of the site of SN~2003gd is presented as Fig. \ref{allimages}.  Broad band photometry of SN~2003gd at multiple late-time epochs is presented as Table \ref{03gdTAB}
  \begin{table}
\caption{\label{03gdTAB}HST and Gemini photometry of SN~2003gd}
\begin{tabular}{lcccc}
Date          &    B      &       V    &      R     &     I    \\ 
\hline
2003 Aug 01   & 19.211    &  17.480    & $\cdots$ &   15.869  \\     
              &(0.002)    & (0.001)    &       &(0.001)\\
\\
\\
2004 Dec 08  &  24.044    & $\cdots$     &  23.152  & $\cdots$ \\
             &(0.037)     &            & (0.041) & \\
\\
\\
2007 June 21 & 25.006     & $\cdots$     & 24.103  & $\cdots$ \\
             & (0.166)    &            & (0.162) &    \\
\\
\\
2007 Aug 11$^{\ast}$  & $\cdots$     & $\cdots$     & 24.445  & 25.270 \\
             &            &            &(0.064) & (0.275)\\
\\
\\
2008 Sep 06/10$^{\dagger}$ & $\cdots$  & 24.997     & 24.651  & $>26.3^{\ddagger}$ \\
               &         & (0.042)    & (0.052) &\\
\end{tabular}
\\
\\
$^{\ast}$ Observed Vega-magnitudes for the F622W(R) and F814W(I) filters in the WFPC2 photometric system.\\
$^{\dagger}$ AB magnitudes in the Sloan $u'g'r'i'z'$ photometric system.\\
$^{\ddagger}$ Assuming $V-I=2$ this corresponds to $I_{J}=25.2$.
\end{table}
\section{Image Subtraction}

\begin{figure}
\hfil~\includegraphics[width=11cm]{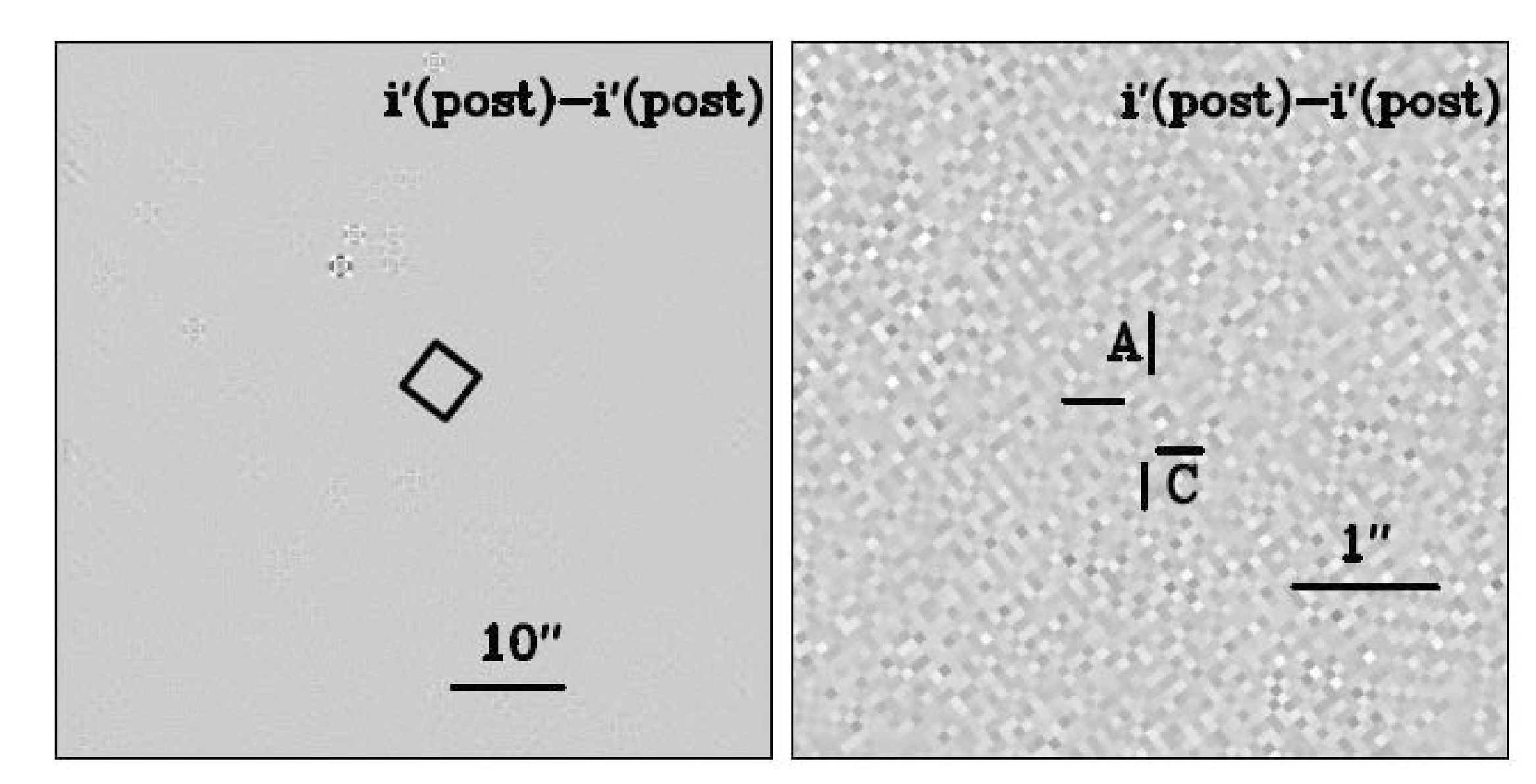}~\hfil
\caption{Self-subtracted post-explosion $i'$ images. {\bf (Left)} $900 \times 900$px frame used in the ISIS package.  The black square indicates the area of the frame in which the SN is located (see next panel).  Minor residuals are associated with extended objects. {\bf (Right)} The area of the self-subtracted frame containing the SN location.  The area covered by the panel corresponds exactly to the panels presented in Fig. \ref{allimages}.}
\label{selfsub}
\end{figure}

Image subtraction, and the photometry of the resulting clean images,
of the SN~2003gd Gemini images was principally conducted using the
{\it ISIS} optimized image subtraction software
\cite{1998ApJ...503..325A,2000A&AS..144..363A}\footnote{http://www2.iap.fr/users/alard/package.html}.
As the 2008 post-explosion Gemini GMOS $i'$ image was acquired under
better seeing conditions than the pre-explosion $i'$ observation of
2001, the 2008 image was aligned with the pre-explosion observation
using the {\it geomap} and {\it geotran} tasks in IRAF. Any
degradation in the seeing in the resulting transformed post-explosion
image was negligible compared to the much larger seeing of the
pre-explosion image.  The images used with the {\it ISIS} package were
smaller $900 \times 900$ stamps extracted from the larger observed
images.  The large area of these stamp images meant there were
sufficient single stars, over a range of brightness, to enable the
construction of a good PSF model. The transformed post-explosion image
was used as the reference frame for subtraction from the pre-explosion
frame using {\it ISIS}.  {\it ISIS} automatically matches the PSF of
the reference image to the input images, as well as appropriately
scaling the flux of objects and the background.  {\it ISIS} produces
subtraction images and photometry for variable objects with different
flux levels in the reference and input images.  Single stellar
objects, confirmed as singular on the post-explosion frame, were
cleanly subtracted on the subtraction image, whereas large extended
objects left minor residuals.\\
The photometry of {\it ISIS} provides a measure of the flux, in
counts, of any variable objects at each epoch, along with a similar
measure of flux for reference objects in the field which were not
variable between the two frames.  Photometry of the progenitor used
the flux level of the progenitor object, reported in counts by {\it
  ISIS}, relative to the flux of these standard stars, which was
converted to a magnitude using comparable {\tt daophot} photometry of
the reference stars on the post-explosion frame and the standard
formula relating magnitudes and relative flux. \\
As {\it ISIS} automatically handles a large number of the details
involved in image subtraction, and is sometimes considered a ``black
box'', a comparable analysis was conducted in parallel using the IRAF
tasks {\it linmatch} and {\it psfmatch}, which scale the flux between
two aligned images and produces a convolution kernel to match the PSF
of one image to another, respectively.  Importantly, having run these
tasks and subtracted the convoluted post-explosion image from
the pre-explosion image an identical result to the {\it ISIS} analysis was achieved.\\
The quality of the PSF and the subtraction process can be estimated by
studying the residual degree of flux in the self-subtracted image
($i'\mathrm{(post)}-i'\mathrm{(post)}$) after convolution and scaling.
At the SN location, the residual flux is consistent with a null
residual [$(0.7\pm44)\,\times10^{-3}$].  The self-subtracted image is shown as Fig. \ref{selfsub}.\\
The background at the SN location in the post-explosion $i'$-band
image is particularly complex (see Fig. \ref{allimages}).  Application
of the {\tt daophot} {\it allstar} routine, with a PSF determined from
the image, does not provided a satisfactory single star solution (for
either the sharpness or $\chi^{2}$ parameters - $\chi^{2}=4$).
Synthetic artificial stars, of known magnitude, were inserted into the
post-explosion frame at the SN position, using the {\tt daophot} task
{\it addstar}, and the {\it ISIS} package was used to recover them.  A
star of magnitude $26.5$ was recovered at an 84\% confidence level
using {\it ISIS}; which we use as an estimate of the detection
threshold at the SN location, on top of the complicated background
structure (as {\it ISIS} only provides a measure of the flux
difference between two images).  We cannot exclude, however, the
possibility that there is still some SN flux that is contributing to
the detected complex background feature.  In this case, therefore, a
more conservative estimate was made using forced photometry at the SN
position, using {\tt daophot} {\it allstar}.  We estimate an upper
limit for the SN+background of $26.3$, which we adopt as a
conservative upper limit on the SN
flux.\\
The likelihood of a unrelated variable object being located at the SN
location was calculated in a simple fashion. The density of objects,
per square pixel, was estimated using all detected variables on the
$900 \times 900$px area.  A more thorough treatment, using
``association probability analysis'', requires appropriate handling of
clustering.  We note that a number of identified variables occur
in dense stellar regions, extended sources or moving objects, which
also leave large residuals in the $i'(post)-i'(post)$ image.  The
measured density is, therefore, an upper limit on the density of true
variable objects, but does not include the effects of ``clustering''
of sources in the vicinity of the SN location (although we note that,
in the case of SN~2003gd, neighbouring stars are resolved and the
residual in the subtraction frame does not have the characteristic
dipole signature of a moving object).  The number of variable objects
randomly expected within the astrometric error circle (see below) is
$0.00001$, suggesting the likelihood of random coincidence to be low.

\section{Differential Astrometry}
The position of the progenitor object of SN~2003gd, on each of the
pre-explosion frames, was determined using differential astrometry.
The position of SN~2003gd was determined from the ACS/HRC F814W
observation at 01 Aug 2003.  Transformations between this image and
the pre-explosion images were calculated using common stars in the
images, and the IRAF task {\it geomap}.  {\it geomap} also provides an
estimate of the r.m.s. uncertainty associated with the transformation.
The SN position from the post-explosion image was transformed to the
coordinates of the pre-explosion images using the IRAF task {\it
  geoxytran}.  {\it ISIS} provides the position of the residual on
the subtraction image.  The difference between the transformed
position and the {\it ISIS} position was $0.023"$, whereas the
uncertainty on the transformation was $0.031"$.  Importantly, for the
analysis presented here, the spatial coincidence is a secondary
concern, since the identity of the progenitor has been confirmed by
its absence in the post-explosion image.

\section{Excess flux correction for ground based imaging of the progenitor of SN~1993J}

\begin{figure}
\hfil~\rotatebox{-90}{\includegraphics[width=6.5cm]{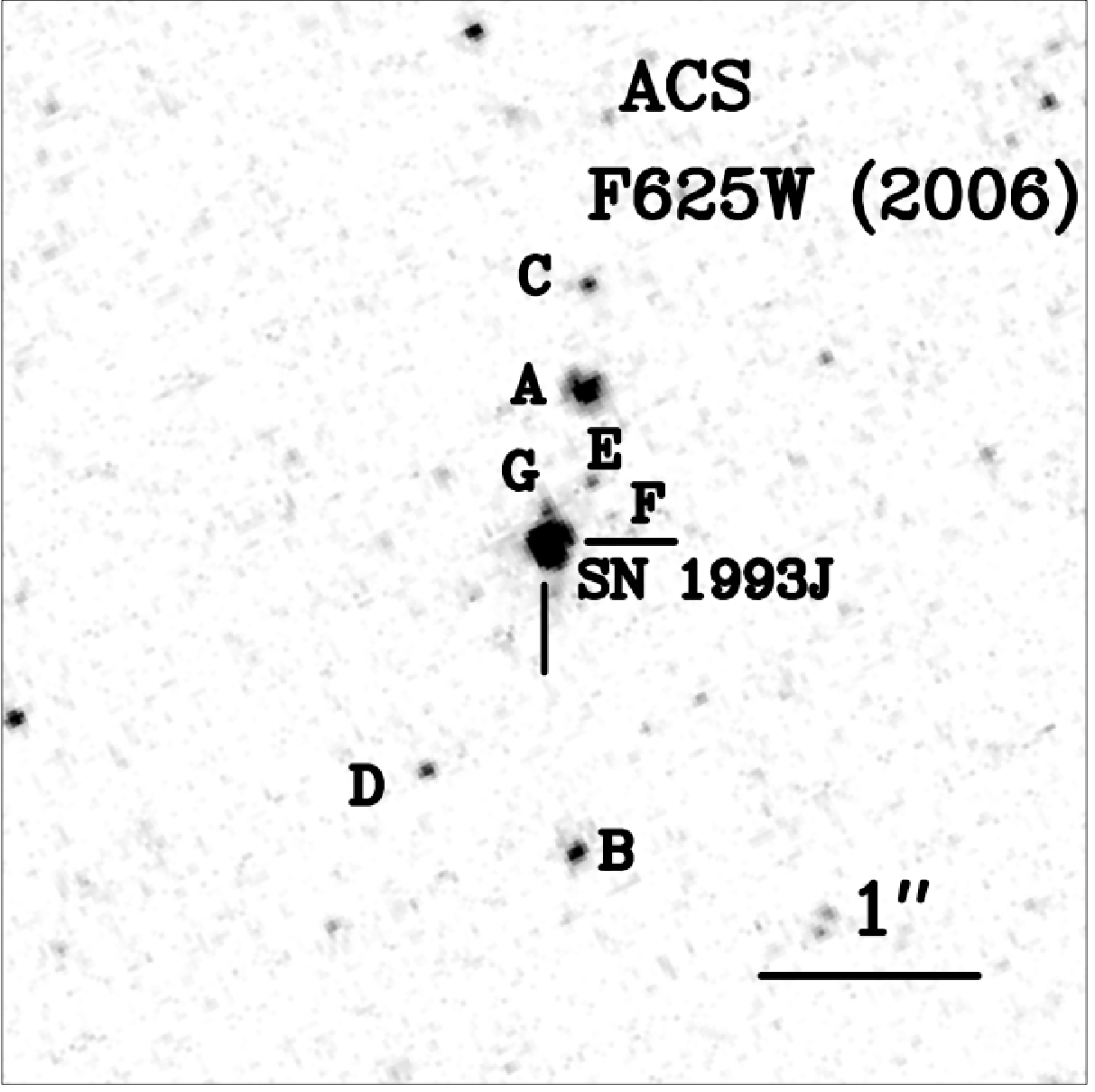}}~\rotatebox{-90}{\includegraphics[width=6.5cm]{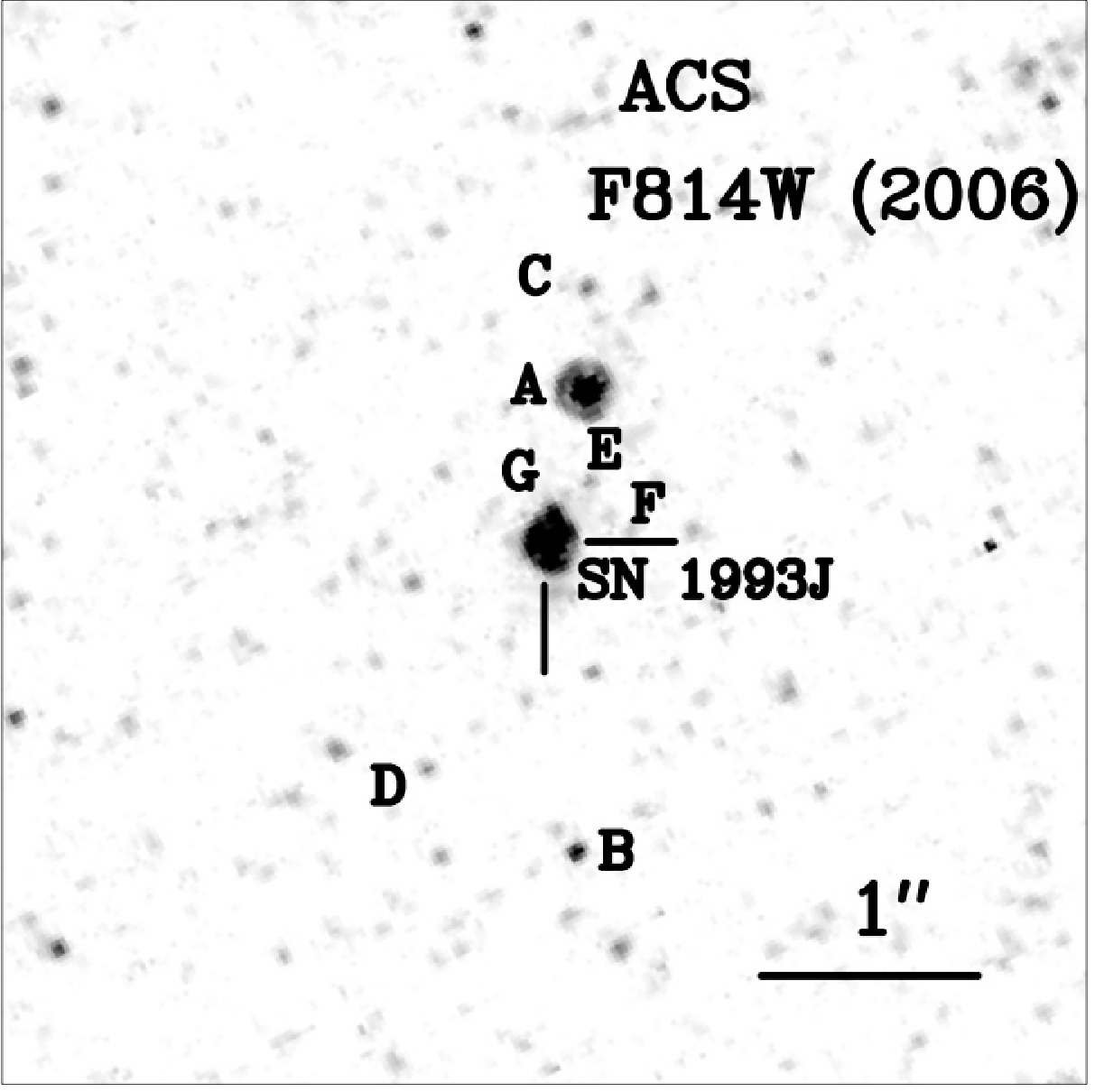}}~\hfil

\caption{The site of SN~1993J.  {\bf Left)} HST ACS HRC F625W image from 03 Nov 2006.  The pixel scale is $0.025"$ per pixel, and the image has been rotated to have North up, East left.  The stars in the vicinity of SN~1993J are labelled according to the established scheme \cite{2002PASP..114.1322V,maund93j}. {\bf Right)} Same as for the left, except an HST ACS HRC F814W image from 01 Nov 2006.}
\label{fig:1993j}
\end{figure}

The site of SN~1993J at 01 Nov 2006 and 03 Nov 2006, in the F814W and
F625W filters respectively, is shown as Fig. \ref{fig:1993j}.  As the
pre-explosion imaging of the progenitor \cite{alder93j} was
ground-based, poor seeing (relative to HST observations) led to a
number of bright nearby stars being unresolved and, hence,
contributing flux to the measured progenitor photometry.  Following
previous work \cite{2002PASP..114.1322V,maund93j}, the excess flux
contribution was determined by using a Gaussian-weighting scheme
(where the flux contribution to the photometry of the progenitor
binary is the total stellar flux for each star scaled by a factor
dependent on the distance of the star from SN~1993J and the seeing,
for each filter, of the pre-explosion observations).  For a general
filter $X$, the excess flux contribution to the progenitor binary photometry
was calculated as:
\begin{equation}
  F_{X}=\sum_{i=A..D}F_{X,i}e^{-r_{i}^{2}/4\sigma_{x}^{2}}
\end{equation} 
where $r$ is the distance in arcseconds of the $i$th star from
SN~1993J (measured from the ACS/HRC images) and $\sigma_{X}$ is the
seeing, for each filter, in the original pre-explosion imaging
\cite{alder93j,2002PASP..114.1322V}.  The pre-explosion photometry is
in the Landolt system, and our ACS photometry has been converted to
the appropriate Johnson $UBV$ and
Cousins $R_{c}I_{c}$ colours \cite{acscoltran}.\\

\begin{figure}
\hfil~\rotatebox{-90}{\includegraphics[width=7cm]{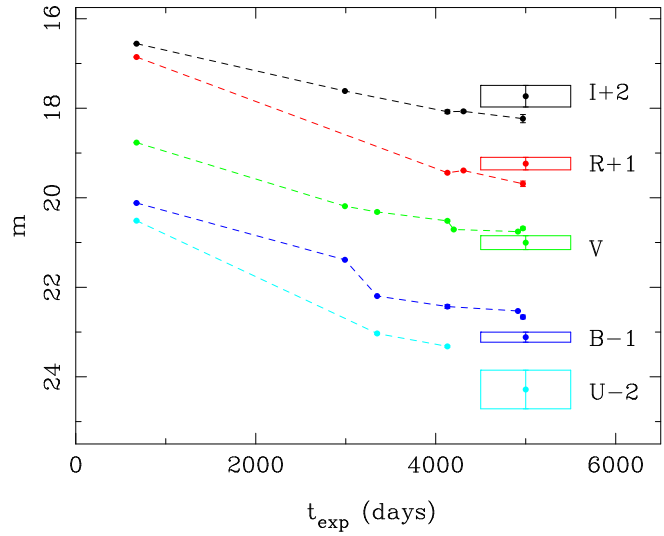}}~\hfil
\caption{The late-time HST light curve of SN~1993J.  The boxes show
  the measured brightness, corrected for excess flux, of the binary
  progenitor system.  The photometric data is plotted in day since the
  date of explosion (JD 2449074; \cite{1994MNRAS.266L..27L}).}
\label{93j:lc}
\end{figure}

The pre-explosion photometry was corrected for the weighted
contributions of stars A, B, C and D (as identified in
Fig. \ref{fig:1993j}), which produced the largest contribution to the
excess flux (in particular in the $U$ and $B$ bands); while Stars E, F
and G were not included in the calculation due to their faintness.
Photometry of these stars in all recent HST WFPC2 and ACS observations
(as listed in Tables \ref{93JACS} and \ref{93JWFPC2}) were used to
determine average $UBVRI$ magnitudes as well as assess photometric
variability.  This photometry is presented in Table \ref{93JEXCESS}.
The subtraction of flux corresponding to a magnitude $m_{2}$ from an
object of magnitude $m_{1}$ gives a new magnitude $m'$ given by:
\begin{equation}
m'=-2.5\log (f_{1} - f_{2} ) =m_{1} - 2.5\log \left (1-\frac{1}{k}\right)
\end{equation}
where $k=10^{(m_{1}-m_{2})/-2.5}$.\\
\begin{table}
\caption{\label{93JEXCESS}Photometry of Stars A, B, C and D in the vicinity of SN~1993J, the excess flux contribution and the corrected progenitor photometry.}
\begin{tabular}{lcccccc}
Star     &       U        &        B      &      V      &     R     &      I     & Distance\\
         &                &               &             &           &            & ($"$)\\
\hline
A$^{\star}$ & 23.387  & 23.808  & 22.634  & 21.523  & 20.530 & 0.74\\
           & (0.017) & (0.362) & (0.878) & (0.644) & (0.300)\\
\\
B   & 23.080 & 23.088 & 22.989 & 23.042 & 22.917 & 1.43\\
    & (0.033)&(0.009)& (0.006) & (0.018) &(0.020) \\
\\
C   &23.079&23.708&23.681&23.742&23.304 & 1.21\\
    &(0.051)&(0.013)&(0.009)&(0.019)&(0.011)\\
\\
D   &22.961&23.704&23.827&23.907&23.906 & 1.19\\
    &(0.065)&(0.014)&(0.011)&(0.023)&(0.033)\\
\hline
$\sigma_{X}$($"$)  & 0.85 & 0.61 & 0.69 & 0.77 & 0.53 \\
\hline
Excess & 22.129 & 23.045 & 22.323 & 21.492 & 20.971 \\
       &(0.024)&(0.124)&(0.495)&(0.450)&(0.277)\\
\hline
AHR$^{1}$ & 21.450 & 21.730 & 20.720 & 19.940 & 19.430 \\
    & 0.200 & 0.070 & 0.040 & 0.080 & 0.170 \\
\hline
AHR Corr.$^{2}$ & 22.283 & 22.114 & 21.002 & 20.237 & 19.731 \\
          &(0.431) & (0.113)&(0.155)&(0.140)&(0.241) \\
\hline
\end{tabular}
\\
$^{\star}$ Star A identified as variable (quoted uncertainty gives combined photometric errors and the range of variability as measured at the epochs of the observations given in Tables \ref{93JACS} and \ref{93JWFPC2}).\\
$^{1}$ Ground-based photometry of the progenitor object, including the excess flux contribution from unresolved nearby stars \cite{alder93j}.\\
$^{2}$ Photometry of the progenitor binary system corrected for the excess flux contribution from nearby stars.\\
\end{table}
The brightness of SN~1993J at 31 Jan 1995, 28 May 2002, 15 Jul 2004 and 1/3 Nov 2006 (as presented in Fig. 2, in the main body of the paper) are given in Table \ref{93JPHOTOM}.\\
The photometry presented of SN~1993J and the surrounding stars are
approximately consistent with previously presented photometry of these
objects \cite{2002PASP..114.1322V,mythesis}.  Minor differences may
have arisen due to intrinsic photometric variability (see Star A),
changes in the adopted zeropoints in the intervening period between
the publication of those studies and differences in the filter
responses between the WFPC2 and ACS instruments.
\begin{table}
\caption{\label{93JPHOTOM}HST WFPC2 and ACS photometry of SN~1993J at 31 Jan 1995, 28 May 2002, 15 Jul 2004 and 1/3 Nov 2006.  Magnitudes are given in the standard Johnson-Cousins $UBVRI$ system.}
\begin{tabular}{lccccc}
Date         &       U        &        B      &      V      &     R     &      I     \\
             &                &               &             &           &            \\
\hline
31 Jan 1995  & 18.512         &  19.116       &   18.767    & 17.856    & 18.557     \\
             & (0.011)        &  (0.006)      &  (0.003)    & (0.002)   & (0.004)    \\ 
\\
28 May 2002  &21.032          & 21.195        &  20.317     &  $\cdots$ & $\cdots$   \\ 
             &(0.024)         & (0.011)       & (0.021)     &           &            \\
\\
15 Jul 2004  & 21.320         & 21.430        & 20.515      &  20.441   & 20.078     \\
             & (0.022)        & (0.042)       & (0.020)     & (0.020)   & 0.041      \\
\\
1/3 Nov 2006 & $\cdots$       & 21.663        & 20.683      & 20.687    & 20.232     \\
             &                & (0.042)       & (0.030)     & (0.062)   & (0.091)    \\
\hline
\end{tabular}
\end{table}
The evolution of the SN~1993J light curve, relative to the limits on the brightness of the binary progenitor system, corrected for the flux excess, is shown as Fig. \ref{93j:lc}.

\bibliographystyle{science}

\end{document}